\documentclass[11pt]{article}

\usepackage{hyperref}
\usepackage{xspace}
\usepackage{times}
\usepackage{cite}
\usepackage{amsmath}
\usepackage[text={7.5in,9.9in},centering]{geometry}
\newcommand{\term}[1]{#1}
\newcommand{\arrow}{\rightarrow}
\newcommand{\stacktext}[2][c] {\begin{tabular}{#1}#2\end{tabular}}
\newcommand{\xmath}[1]{\ensuremath{#1}\xspace}
\newcommand{\x}{\xmath{x}}
\newcommand{\y}{\xmath{y}}
\newcommand{\xh}{\xmath{\hat{x}}}
\newcommand{\Erdogan} {Erdo\u{g}an}
\newcommand{\Boernert} {B\"ornert}
\newcommand{\Candes} {Cand\`es}
\long\def\comment#1{}

\begin{document}

\textbf{%
Medical image reconstruction:
a brief overview of past milestones
and future directions
}
\\
Jeffrey A. Fessler
\\
University of Michigan
\\
\today

\textsl{

At ICASSP 2017,
I participated in a panel
on ``Open Problems in Signal Processing'' 
led by Yonina Eldar and Alfred Hero.
Afterwards the editors of
the IEEE Signal Processing Magazine
asked us to write a ``perspectives'' column
on this topic.
I prepared the text below
but later found out that equations or citations
are not the norm in such columns.
Because I had already gone to the trouble
to draft this version
with citations,
I decided to post it on arXiv in case
it is useful for others.

}


Medical \term{image reconstruction}
is the process of forming interpretable images
from the raw data recorded by an imaging system.
Image reconstruction is an important example
of an \term{inverse problem}
where one wants to determine the input to a system
given the system output.
The following diagram illustrates the data flow
in an medical imaging system.
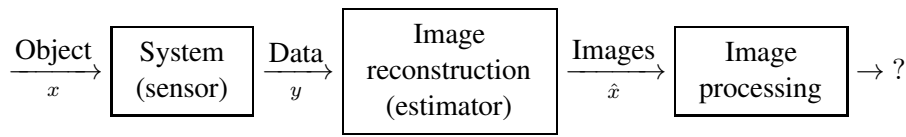
\[
\xrightarrow[\x]{\mbox{Object}}
\fbox{\stacktext{System \\ (sensor)}}
\xrightarrow[\y]{\mbox{Data}}
\fbox{\stacktext{Image \\ reconstruction \\ (estimator)}}
\xrightarrow[\xh]{\mbox{Images}}
\fbox{\stacktext{Image \\ processing}}
\arrow \ ?
\]

Until recently,
there have been two primary methods
for image reconstruction:
\term{analytical} and \term{iterative}.
Analytical methods
for image reconstruction
use idealized mathematical models
for the imaging system.
Classical examples are the filtered back-projection method
for tomography
\cite{cormack:63:roa,shepp:74:tfr,fessler:14:foc}
and the inverse Fourier transform
used in magnetic resonance imaging (MRI)
\cite{wright:97:mri}.
Typically these methods consider
only the geometry and sampling properties of the imaging system,
and ignore the details of the system physics
and measurement noise.
These reconstruction methods have been used extensively
because they require modest computation.
Despite the long history of such methods,
interesting advances continue to made,
particularly for incomplete data
\cite{clackdoyle:10:tri}.

Over the past two decades,
image reconstruction
has evolved
from exclusive use of analytical methods
to wider use of
iterative or
\term{model-based} methods
that account for the physics of the imaging system
and the statistical properties of the measurement noise
\cite{
ollinger:97:pet,
lewitt:03:oom,
qi:06:irt,
fessler:10:mbi,
nuyts:13:mtp%
}.
These properties are captured by a likelihood function
$p(\y|\x)$
and accurate modeling
requires thorough understanding of an imaging system.
Usually the number of unknown voxel values in \x
is comparable to the number of measurements in \y
(or even fewer)
so the problems are under-determined
or poorly conditioned
so \term{maximum-likelihood} (ML) methods
would propagate excessive noise
from the measurements into the reconstructed image \xh.
Using priors $p(\x)$
or regularizers
can overcome this limitation,
so most iterative methods
used for image reconstruction
have been based on
\term{maximum a posteriori} (MAP) estimation
by seeking the maximizer of the posterior
$p(\x|\y)$
or equivalently
(by Bayes rule)
the sum of the log likelihood
and the log prior
\begin{equation}
\xh = \arg \max_{\x} \log p(\y|\x) + \log p(\x)
.\end{equation}
This equation captures most of the research topics
in image reconstruction:
(i)
modeling the system physics and statistics
in the likelihood;
(ii)
developing signal models to serve as priors;
(iii)
developing faster optimization algorithms;
and
(iv)
assessing the quality of the reconstructed image \xh.
The signal processing community
has had a particularly important role
in developing signal models.
Numerous signal models have been explored
for image reconstruction over the years,
such as Markov random fields and wavelets.
The optimization methods used in medical imaging
also have much in common
with methods used in machine learning
and other large-scale problems.

The transition from analytical to iterative methods
took place at widely different dates
in different modalities.
In PET and SPECT,
a seminal paper
on an \term{expectation maximization} (EM) algorithm
in the early 1980's
\cite{shepp:82:mlr,lange:84:era}
led to over a decade of research
before a key acceleration method
called \term{ordered subsets} (OS)
\cite{hudson:94:air}
(related to \term{incremental gradients}
in the optimization field)
helped lead to commercial adoption
of OS-EM for clinical PET and SPECT
in about 1997,
using an (unregularized) ML approach.
This transition provided a dramatic improvement in image quality
because PET data is very noisy
so modeling the statistics is crucial.
For years the human PET scanners used
unregularized ML methods
while some animal scanners
had a more sophisticated MAP method
\cite{qi:98:hr3}!
Human PET scanners only recently
began to provide MAP methods clinically
\cite{ahn:15:qco}
using a modification of a Gaussian MRF prior
\cite{nuyts:02:acp}
and a convergent OS algorithm
\cite{ahn:03:gci}.

In X-ray CT,
iterative image reconstruction
first became available commercially
for the CT part of SPECT-CT scanners
in about 2010,
\cite{hansis:10:irf},
using a different OS algorithm
published a decade earlier
\cite{erdogan:99:osa}.
In 2012,
the first FDA-approved iterative MAP method
targeted at reduced X-ray dose
(the primary impetus for iterative methods in CT)
became available for clinical CT
\cite{thibault:07:atd},
building on a
publication in the signal processing transactions
from two decades earlier
\cite{sauer:92:beo}.
This method also uses a modified
Gaussian MRF to make it edge-preserving.

In MRI,
iterative methods were introduced in research labs
to quantify relaxation parameters
\cite{miller:95:mbm},
reconstruct data from multiple receive coils
\cite{pruessmann:01:ais},
or correct for magnetic field inhomogeneities
\cite{sutton:03:fii},
among other considerations.
None of these methods have been adopted clinically.
However,
a key turning point in the field
was the introduction of compressed sensing
in about 2005
\cite{candes:05:srf,candes:06:rup,donoho:06:cs,%
baraniuk:07:cs,candes:08:ait,romberg:08:ivc}
and its rapid illustration
on real MRI applications in about 2007
\cite{lustig:07:smt,lustig:08:csm}.
This led to an explosion of research
that finally led to FDA approval
of compressed sensing MRI products
in 2017 by two major MRI vendors
\cite{fda:17:ge,fda:17:siemens}
with others soon to follow.
Combinations of \term{total variation} regularization
and wavelet sparsifying transforms
are used widely in this field.

In all the above examples,
over a decade passed
between the key publication
and commercial availability
of the method!
New methods typically require too much computation
to be practical immediately.
The importance of ensuring that new methods
lead to comparable or improved diagnoses
means that considerable investigation is needed.

All of the above MAP methods available commercially
for image reconstruction
use relative simple regularizers (priors)
defined mathematically
such as MRFs and wavelets.
The emerging trend in the field
is to replace human-defined signal models
with signal models that are
\emph{learned} from data.
For example in X-ray CT,
there are numerous CT images already acquired
at ``normal'' X-ray doses,
and one can learn signal models
such as dictionaries
from that training data
and then use those signal models later
to reconstruct images
from low-dose or limited-view data
\cite{lu:12:fvi,xu:12:ldx}.
Another data-driven option is to learn a sparse signal model
concurrently with the image reconstruction process,
rather than relying on prior training data.
This approach is known as \term{blind} or \term{adaptive}
dictionary (or transform) learning
\cite{ravishankar:11:mir,ravishankar:15:ebc,ravishankar:16:ddl}.
These methods are a fairly radical departure
from the previous 3+ decades of image reconstruction research
where most regularizers were defined using math models and physics,
not from data.
This evolution provides opportunities
for signal processing researchers
to explore data-driven signal models
to better solve inverse problems,
particularly from limited or noisy data.
For multi-dimensional data
such as spectral CT or dynamic MRI,
tensor models are of growing interest
\cite{yu:14:mcs,zhang:17:tbd}.

In addition to learning a regularizer,
one can ``unroll the loop''
of an iterative algorithm for image reconstruction
and think of the resulting block diagram
as a sequence of processing steps
akin to a \term{deep neural network}
and then use data to train more aspects
of the processing chain.
The earliest such unrolling
was probably \term{learned ISTA} (LISTA)
\cite{gregor:10:lfa}
and recent conferences
have seen an explosion of methods of this kind
\cite{wang:16:apo}.
Neural nets are often trained using stochastic gradient descent
of a simple loss function like mean-squared error,
but these metrics may not be the most meaninful
for medical imaging applications.
Training with loss functions related to image quality
\cite{zhao:17:lff}
(and ultimately diagnostic performance)
is needed.
There are many additional significant challenges.
These methods are arguably even more nonlinear
than the edge preserving regularization methods
used clinically today (in CT for example).
Can one characterize
the ``resolution'' and ``noise'' properties
of such methods?
What is the best training metric - MSE or diagnostic image quality?
What if a patient has significantly different image features
than those found in the training data?
Will methods trained on, say, diffusion brain MRI scans
generalize to T1-weighted cardiac perfusion scans?
MRI in particular has so many different types of image constrast
that it would seem to require an enormous amount of training data
to cover all cases.
How well will a method trained for one system configuration
(e.g., a certain set of coils in MRI
or a certain set of angular views and pitch in CT)
generalize to other configurations?
Some experts have conjectured that
``machine learning will transform radiology significantly
within the next five years''
\cite{wang:17:mlw},
but others point out there are significant technical and legal challenges.
These questions and more
should provide numerous research opportunities
for signal processors interested in inverse problems
like medical imaging.

\newpage



\end{document}